\documentclass[twocolumn,showpacs,preprintnumbers,amsmath,amssymb]{revtex4}

\usepackage{graphicx}
\usepackage{dcolumn}
\usepackage{bm}
\usepackage{subfigure}

\begin{document}


\title{Guiding chemical pulses through geometry: Y-junctions}

\author{L. Qiao}
\author{I. G. Kevrekidis}
\altaffiliation[Also at ]{the Program in Applied and Computational
Mathematics(PACM) , Princeton University, Princeton,NJ 08544, USA}
\affiliation{Department of Chemical Engineering, Princeton
University, Princeton, NJ 08544 USA}
\author{C. Punckt}
\author{H. H. Rotermund}
\affiliation{Fritz-Haber-Institut der MPG, Faradyweg 4-6, 14195
Berlin, Germany}

\date{\today}

\begin{abstract}
We study computationally and experimentally the propagation of
chemical pulses in complex geometries.
The reaction of interest, CO oxidation, takes place on single
crystal Pt(110) surfaces that are microlithographically patterned;
they are also {\it addressable} through a focused laser beam,
manipulated through galvanometer mirrors, capable of locally
altering the crystal temperature and thus affecting pulse
propagation.
We focus on sudden changes in the domain shape (corners in a
Y-junction geometry) that can affect the pulse dynamics; we also
show how brief, localized temperature perturbations can be used to
control reactive pulse propagation.
The computational results are corroborated through experimental
studies in which the pulses are visualized using Reflection
Anisotropy Microscopy.
\end{abstract}

\pacs{82.40.Ck, 82.40.Np, 05.45.-a}
\maketitle

\section{Introduction}
\label{intro}

Reaction control is an essential task in the chemical process
industry; it can stabilize unstable (but profitable) operating
states, and, in a multiobjective context, can enhance reactor
performance while satisfying safety and environmental constraints.
Non-steady state (e.g. periodic) operation may result in better
average productivity or selectivity than steady state operation
(as suggested in the early 1960s in the pioneering work of F. Horn
(see Ref.~\cite{n-s-reactor} and references therein)).
At a much finer scale, the nonlinear characteristics of
heterogeneous catalytic reactions on single-crystal catalysts
could be exploited to improve overall yield and selectivity
through spatiotemporally resolved actuation
\cite{addressing,twists}.
Micro-lithography, by designing the shape of reactive domains as
well as prescribing the geometry and statistics of heterogeneous
inclusions, provides a different avenue of ``talking" to the local
dynamics of catalytic reactions (e.g. Ref.~\cite{chaos}).
The phenomenology of pattern formation on micro-composite
catalytic surfaces can be much richer than that in uniform media.
There have been extensive experimental and computational studies
in this area, both from our group
\cite{pwork1,pwork2,pwork3,pwork4,pwork5,pwork6,pwork7,pwork8,pwork9,pwork10}
and in the work of others \cite{pwork11,pwork12,pwork13}.
For the low pressure CO oxidation on Pt(110), the heterogeneities
deposited onto the catalytic Pt surface can be either inert, such
as TiO$_2$, or may consist of different active catalysts for the
reaction, e.g. metals like Rh and Pd.
Such fields of heterogeneous inclusions affect the reaction
dynamics and the formation of patterns on the catalytic Pt surface
largely through their interfaces.


The heterogeneity in such composite catalysts is solely spatial;
their geometry is determined upon construction and does not change
with time.
In more recent studies, spatiotemporal forcing (using a
galvanometer mirror-manipulated focused laser beam) was applied
during CO oxidation on a Pt(110) single-crystal \cite{addressing}.
Pulses and fronts, the basic building blocks of patterns, could be
formed, guided and destroyed with an addressable laser beam; the
addressability function was also used to enhance catalytic
reactivity in Ref.~\cite{twists}.

From the implementation of chemical ``logical gates" (e.g.
Ref.~\cite{Showalter}) to the recent spatiotemporal control of
morphogenetic patterns in drosophila embryos \cite{Lucchetta}, and
from drop formation control in microfluidics \cite{Joanicot} to
the guidance of matter waves in Bose-Einstein condensates
\cite{PKEV}, guiding pulses and fronts in complex geometries is an
essential task in spatiotemporal pattern control.
What makes it increasingly possible is the combination of
spatiotemporally finely resolved sensing combined with
spatiotemporally finely resolved actuation.
Whether the pattern-forming system is naturally photosensitive,
like the chemical waves in Ref.~\cite{Sakurai} or rendered
photosensitive, like the neurons in Ref.~\cite{Lima} through
genetic manipulation, rapid technological developments in finely
resolved optical actuation techniques (e.g. see Ref.~\cite{Grier})
are rapidly and radically changing the experimental exploration
and control of spatiotemporal pattern formation.

In this paper we explore the combined effects of geometry and of
spatiotemporally resolved addressability.
In particular, we explore the influence of inert boundary geometry
on the dynamics of propagating pulses for CO oxidation on a
micro-composite Pt(110) surface, and show how these dynamics can
be altered using an addressable laser beam.
Our geometry of choice for pulse propagation studies is a Y-shaped
junction structure, in which sharp boundary curvature changes
(corners) can effectively dictate pulse propagation.
We systematically explore the effect of varying the junction
geometry on reactive pulse propagation; a related rhomb geometry
is also studied with qualitatively similar results.
We then show how one can {\it actively} alter the phenomena
dictated by geometry through the use of single-shot
spatiotemporally localized laser heating.
The computational predictions are validated by experimental
observations of reactive pulses visualized through Reflection
Anisotropy Microscopy (RAM) \cite{RAM}.

\section{Modeling}

In our simulations we use the three-variable
Krischer-Eiswirth-Ertl reaction-diffusion model for CO oxidation
on Pt(110) with a surface phase transition described in
Ref.~\cite{model}.
This surface reaction follows a Langmuir-Hinshelwood mechanism:

\[CO+* \leftrightharpoons CO_{ads}\]
\[2*+O_2 \rightarrow 2O_{ads}\]
\[CO_{ads}+O_{ads} \rightarrow 2*+CO_2\uparrow
\]
accompanied by a $1\times2\rightarrow1\times1$ phase transition of
the Pt(110) surface due to CO adsorption.
When the coverage of CO lies between 0.2 and 0.5, the fraction of
$1\times1$ surface increases monotonically as the CO coverage
increases.
The sticking coefficient of oxygen is $50\%$ higher on the
$1\times1$ surface as compared to the $1\times2$ surface.
The $1\times1$ phase favors oxygen adsorption, which leads to
reactive consumption of CO.
This can lead to oscillatory behavior of the reaction, and also
allows the formation of propagating reaction pulses.

The equations for this kinetic model are

\[\dot{u}=k_us_up_{co}\left(1-\left(\frac{u}{u_s}\right)^3\right)-k_1u-k_2uv+D_u\nabla^2u\]
\[\dot{v}=k_vp_{o_2}(ws_{v_1}+(1-w)s_{v_2})\left(1-\frac{u}{u_s}-\frac{v}{v_s}\right)^2-k_2uv\]
\[\dot{w}=k_3(f(u)-w)
\]
where by $u$, $v$ and $w$ we denote the surface coverage of CO
and O, and the surface fraction of $1\times1$ phase respectively.
The adsorption rate constants for CO and O$_2$, $k_u$ and $k_v$
respectively, are considered to be constant within the temperature
range considered here.
The rate constants $k_1,k_2$ and $k_3$ for the desorption,
reaction and surface phase transition are given by the Arrhenius
formula $k_i=(k^0)_iexp(-E_i/RT)$ and T is the temperature of the
single-crystal.
The function $f(u)$ is fit to experimental data to give the rate
of surface phase transition as a function of $u$, the CO surface
coverage, as follows:

\[f(u)=\left\{\begin{array}{ccc}0&\mbox{for}&u\leq0.2\\
\frac{u^3-1.05u^2+0.3u-0.026}{-0.0135}&\mbox{for}&0.2<u<0.5\\
1&\mbox{for}&u\geq0.5\end{array}\right.\]

For simplicity, the diffusion of CO is taken to be isotropic and
no-flux boundary conditions are used in our simulations.
To reflect the influence of laser heating on the dynamics of the
reaction, we approximate the local temperature increase caused by
a laser spot through a local Gaussian temperature field.
The heat generated by the reaction ($\sim1mW$) can be neglected
compared to the power of the laser beam($\sim1W$).
Since the diffusivity of adsorbed CO ($\sim10^{-8}cm^2/s$) is much
smaller than the thermal diffusivity constant of Pt
($\sim10^{-1}cm^2/s$), we can assume the local Gaussian
temperature field is established (resp. vanishes) instantaneously as the
laser beam is applied (resp. removed) \cite{dragging,Cisternas}.
In our simulations, we use the commercial Finite Element package
FEMLAB to compute the time-dependent evolution of reactive pulses
in 2D Y-junction geometries; our meshes typically consisted of
$\sim 12,000$ linear elements.

\section{Computational results}

\subsection{Pulse propagation in a Y-junction}

\begin{figure}
\centering
\includegraphics[width=0.9\columnwidth]{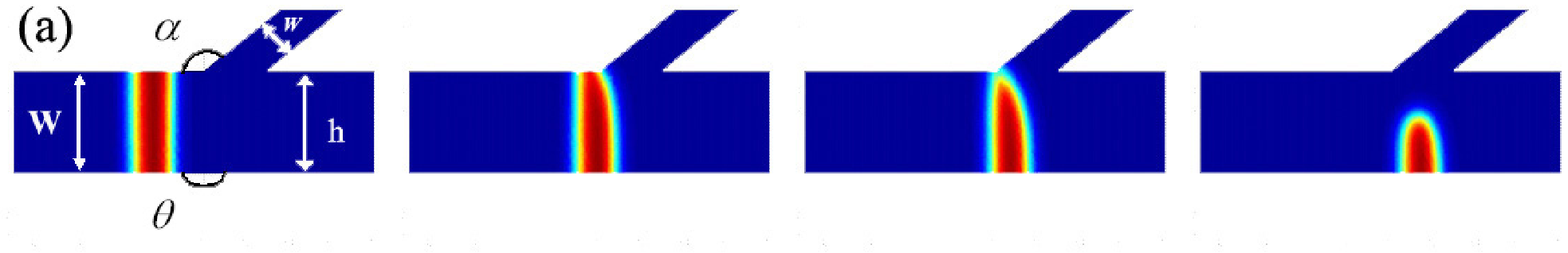}
\includegraphics[width=0.9\columnwidth]{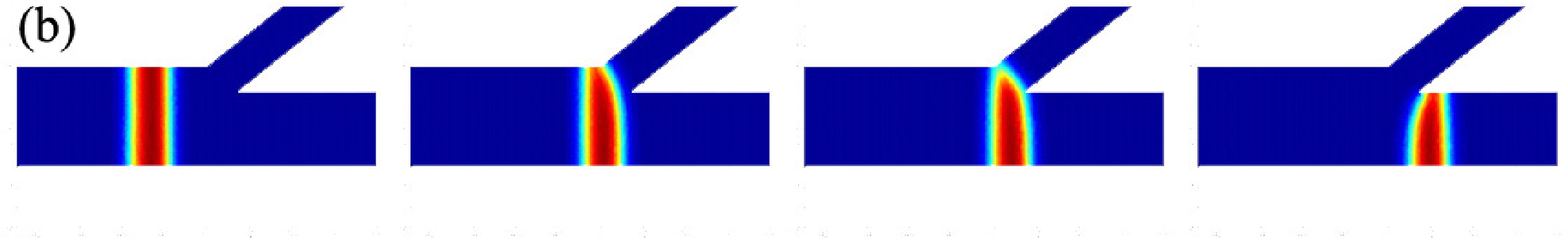}
\includegraphics[width=0.9\columnwidth]{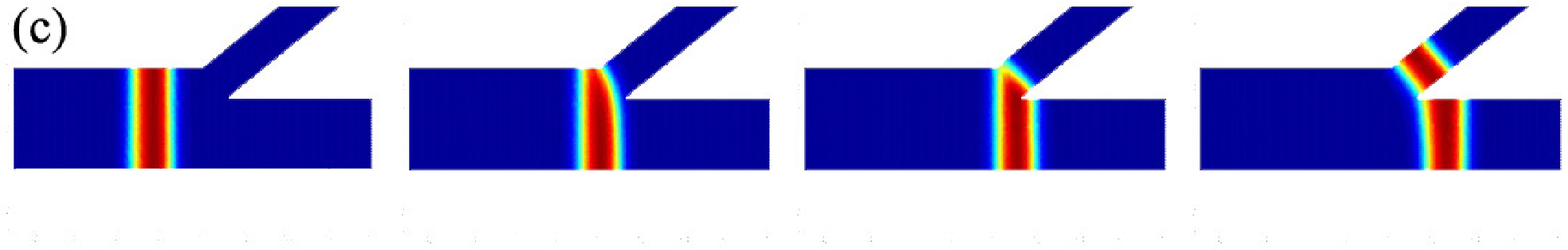}
\includegraphics[width=0.9\columnwidth]{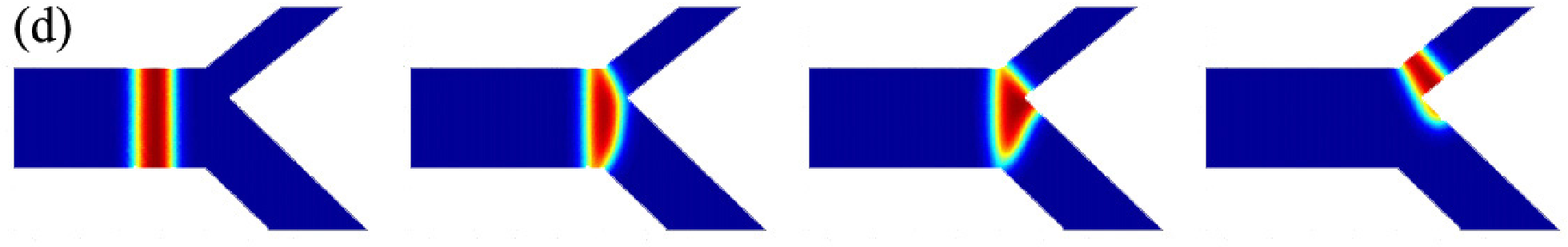}
\caption{(Color online) Pulse propagation in different Y-junction
structures.
By choosing appropriate geometric parameters, we are able to
dictate the pulse transmission patterns.
(a) $h=5$, $\theta=\pi$, (b) $h=3.7$, $\theta=\pi$, (c) $h=3.5$,
$\theta=\pi$, (d) $h=3.5$, $\theta=\frac{3}{4}\pi$.
Other parameters: $W=5$, $w=1.93$, $\alpha=\frac{7}{9}\pi$,
$T=535.5K$, $P_{CO}=4.95\times10^{-5}mbar$,
$P_{O_2}=2.0\times10^{-4}mbar$.} \label{y_propagation}
\end{figure}

The geometry of the Y-junction structure is described through the
parameters $W$, $w$, $h$, $\alpha$, and $\theta$ (see the first
snapshot in Fig.~\ref{y_propagation}(a)).
When a reactive pulse reaches the corners of the Y-junction that
are denoted by the two angles $\alpha$ and $\theta$, the pulse
front starts becoming convex.
The local travelling speed of the pulse decreases due to this
convex curvature \cite{K_v1,K_v2}.
When the local curvature required to ``go around" one of the
corners exceeds a certain limit, the pulse loses stability,
``decollates" from the boundary, and disappears \cite{K_v3}.
With appropriate angles $\alpha$ and $\theta$, by adjusting the
position of the ``prow" to the exit (i.e. adjusting $h$), we can
force the pulse to choose none, one, or both of the channels to
propagate in (Fig.~\ref{y_propagation}).
The behavior of the pulse also depends strongly on the selected
partial pressures of CO and O$_2$.
Under reaction conditions for Fig.~\ref{y_propagation}, decreasing
the amount of CO in the gas phase stabilizes the pulse.
Below some critical value for $P_{CO}$ the pulses will propagate
through both channels, essentially independent of the selected
geometry.

\subsection{Quantifying the geometry-induced instability in
reactive pulse propagation}

\begin{figure}
\centering
\includegraphics[width=0.6\columnwidth]{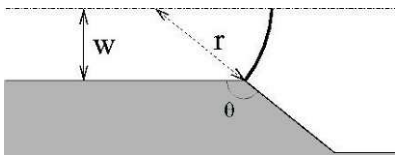}
\caption{Geometry of a corner structure,
$r=\frac{W}{sin(\pi-\theta)}$.} \label{y_structure}
\end{figure}

\begin{figure}
\centering
\includegraphics[width=0.9\columnwidth]{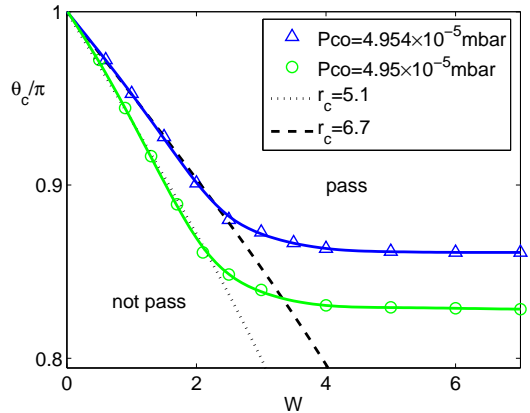}
\caption{(Color online) Critical angle $\theta_c$  for a pulse to
turn around a corner, as a function of dimensionless channel width
$W$.
For fixed CO pressure, pulses in a geometry with $\theta$ and $W$
chosen ``above" the solid lines shown are able to turn around the
corner.
The solid lines, fitted to data points, are included to guide the
eye.
The dashed and dotted lines are given by
$r_c=\frac{W}{sin(\pi-\theta_c)}$ for different $r_c$.
The unit length of $W$ corresponds to a real length of
3.7${\mu}m$.
$T=535.5K$, $P_{O_2}=2.0\times10^{-4}mbar$.} \label{Y_Pco}
\end{figure}

%
We now begin a quantitative exploration of the geometry-induced
instability of pulse propagation in our model system.
Junctions between different linear ``corridors" are an important
building block for complex geometries.
The two geometrical parameters characterizing a junction are the
width of the channel $W$ and the angle $\theta$ at the corner
(Fig.~\ref{y_structure}).
When the reaction conditions are fixed, depending on the choice of
$W$ and $\theta$ , the pulse may or may not be able to pass the
corner, as shown in Fig.~\ref{Y_Pco}.

When a pulse attempts to turn around a corner, if $W$ is small,
the entire pulse becomes curved as a circular arc due to the
influence of the no-flux boundaries, and the curvature of the
pulse is given by $\frac{1}{r}=\frac{sin(\pi-\theta)}{W}$.
If this geometry-determined curvature becomes larger than a
critical value $\frac{1}{r_c}$, a constant determined by the
dynamics of the system and the reaction conditions, the pulse
becomes unstable and disappears~\cite{K_v3}.
This disappearance occurs dynamically through a ``decollation"
mechanism \cite{2Dring}.
When $W$ is large, the pulse only curves locally, close to the
corner, and further increase in $W$ has little influence on this
high local curvature of the pulse.
When $W$ is above some critical value (about 6 for curves in
Fig.~\ref{Y_Pco}), this local curvature of the pulse reaches a
minimum and becomes independent of $W$ and only a function of
$\theta$.
If this minimum value of local curvature is still larger than
$\frac{1}{r_c}$, the pulse will decollate from the corner and fail
to pass through.
We can thus rationalize the existence, for each given set of
reaction conditions, of a critical angle below which the pulse can
not propagate around the corner for any $W$.
%

\begin{figure}
\centering
\includegraphics[width=0.6\columnwidth]{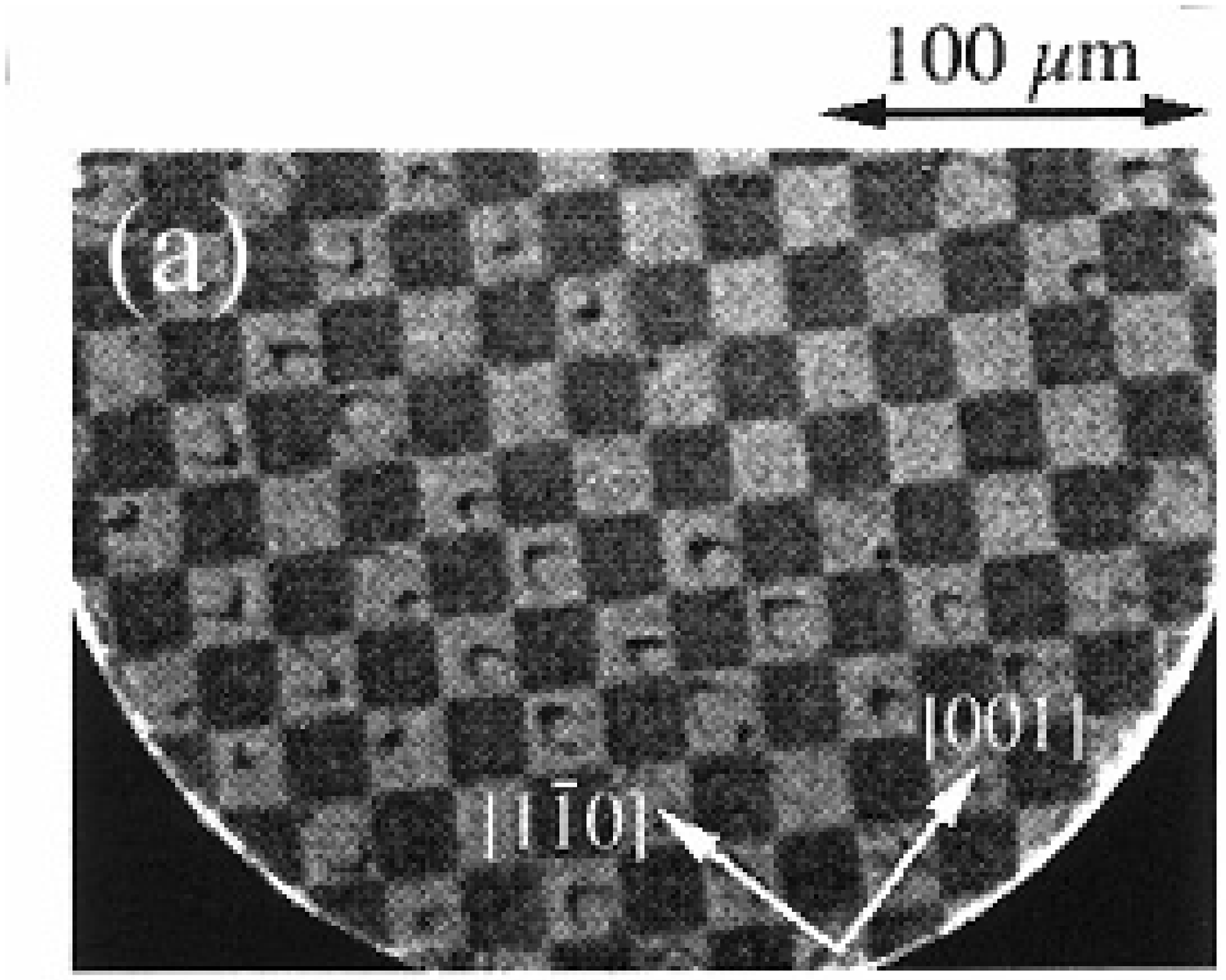}
\includegraphics[width=0.9\columnwidth]{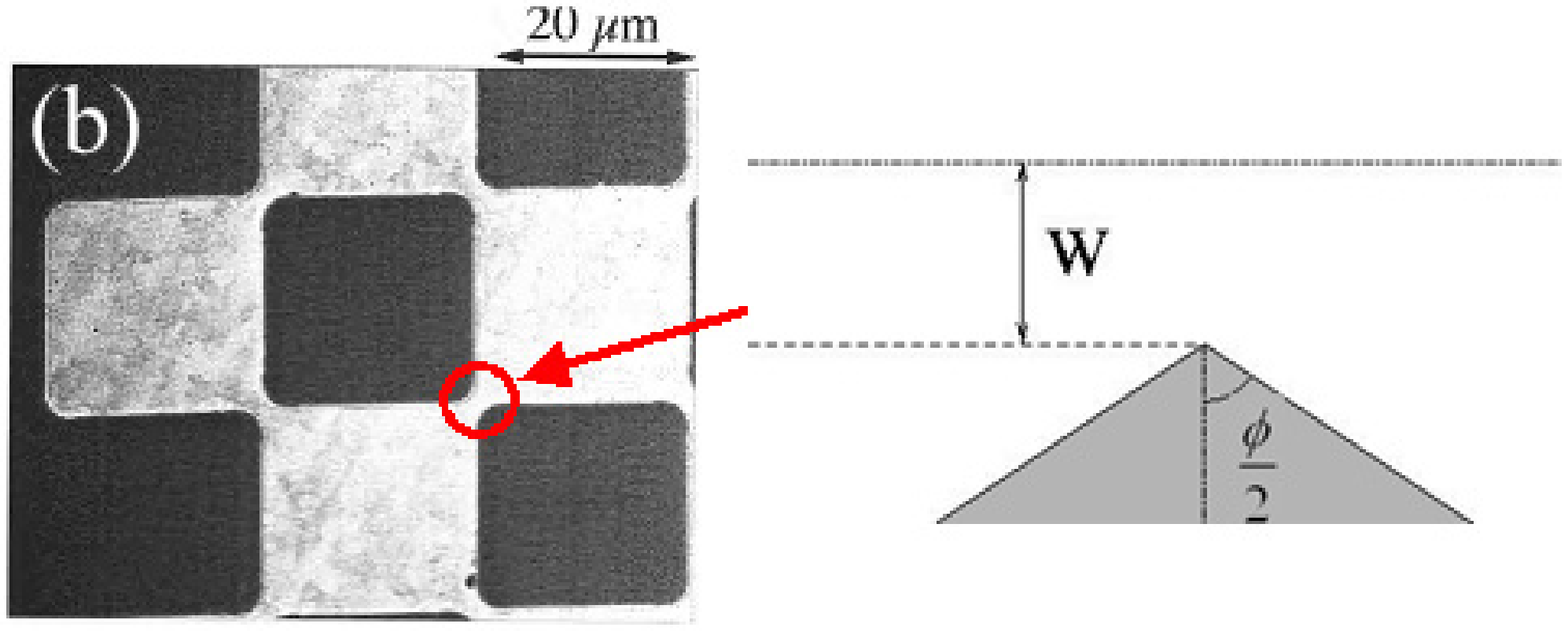}
\caption{(a) A snapshot showing a microdesigned TiO$_2$/Pt
checkerboard.
The inactive TiO$_2$ rhombs appear black while pulses (little arcs
in the first snapshot) are propagating on the light-gray Pt(110)
surface.
(b) A blow-up of the checkerboard structure.
To its right, a geometry (symmetric around the centerline) we used
in the simulations of the rhomb constriction is depicted.}
\label{rhomb}
\end{figure}

Another geometric structure with features similar to the
Y-junction is the rhomb constriction, which arises in
microcomposite TiO$_2$/Pt checkerboards (Fig.~\ref{rhomb}).

\begin{figure}
\centering
\includegraphics[width=0.9\columnwidth]{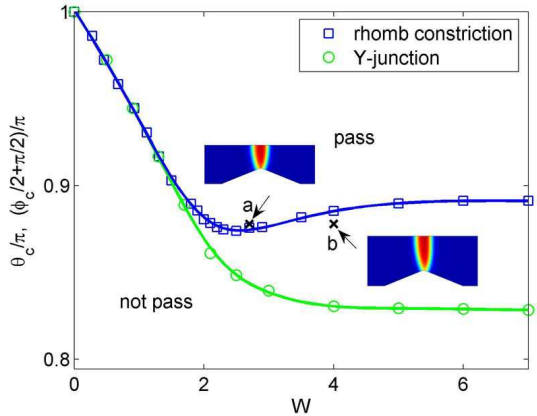}
\caption{(Color online) Critical angle for pulse turning around
the corner as a function of channel width $W$ for the rhomb
constriction and the Y-junction structure.
For the rhomb constriction structure, we plot $\phi_c/2+\pi/2$ vs
$W$  instead of $\phi_c$ vs $W$.
For rhomb constriction (Y-junction) structures with values of $W$
and $\phi/2+\pi/2$ ($\theta$) chosen ``above" the corresponding
curve, pulses are able to turn around the corner.
The solid lines, fitted to data points, are included to guide the
eye.
Snapshots of pulse propagating in the rhomb constriction structure
at point $a$ ($2.7$, $0.88$) and $b$ ($4.0$, $0.88$) are shown in
the insets.
$T=535.5K$, $P_{CO}=4.95\times10^{-5}mbar$,
$P_{O_2}=2.0\times10^{-4}mbar$.} \label{Y_rhomb_snapshots}
\end{figure}

The critical angles $\theta_c$ (for the Y-junction) and $\phi_c$
(for the rhomb constriction) that allow the pulse to propagate
through the structure are plotted as a function of the width $W$,
as in Fig.~\ref{Y_rhomb_snapshots}, under the same reaction
conditions.
The y-axis in Fig.~\ref{Y_rhomb_snapshots} for the rhomb
constriction structure is chosen to be $(\phi/2 + \pi/2)/\pi$, so
that the two critical curves correspond (they are almost
identical) when $W$ is small ($<1$).
The non-monotonic decrease of $\phi_c$ for the rhomb constriction
structure can be rationalized by a simple geometric observation:
as $W$ increases, the front and back of the pulse are approaching
closer and closer to each other at the corner (see the insets in
Fig.~\ref{Y_rhomb_snapshots}); this can lead to a different
mechanism for the decollation of the pulse.

\subsection{Pulse manipulation with local laser heating}

\subsubsection{Using the laser to assist pulse propagation}

The local increase of temperature caused by a short laser ``burst"
accelerates the local desorption of CO and may thus assist the
propagation of an oxygen pulse.
We can therefore use local laser heating to assist pulse
propagation around corners that would be too sharp for the pulse
to go through at isothermal conditions.
\begin{figure}
\centering
\includegraphics[width=0.9\columnwidth]{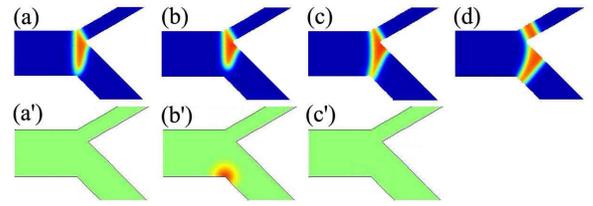}
\caption{(Color online) Using local laser heating to assist pulse
propagation around corners.
Snapshots (a)-(d) show the coverage of oxygen; corresponding
instantaneous temperature fields before, after ((a') and (c')
respectively) as well as during the laser shot ((b')) are plotted
in the second row.
Laser heating is turned on for a total of 0.4s (from t=5.0s to
5.4s) and centered at the lower junction corner; the maximum
temperature increase there is $3K$.
(a) t=4.5s, (b) t=5.0s, (c) t=5.5s, (d) t=8.0s.
$T=535.5K$, $P_{CO}=4.95\times10^{-5}mbar$,
$P_{O_2}=2.0\times10^{-4}mbar$. } \label{laser_help}
\end{figure}

In Fig.~\ref{laser_help}, a local increase of temperature (see the
high temperature field close to corner in
Fig.~\ref{laser_help}(b')) reignites the CO-poisoned surface, and
reattaches the decollated oxygen pulse back towards the corner
(close to the corner in Figs.~\ref{laser_help}(c) and
\ref{laser_help}(d)).
This result has been confirmed experimentally as we will see
below.

\subsubsection{Using the laser to prevent pulse propagation}

\begin{figure}[htp]
\centering
\includegraphics[width=0.9\columnwidth]{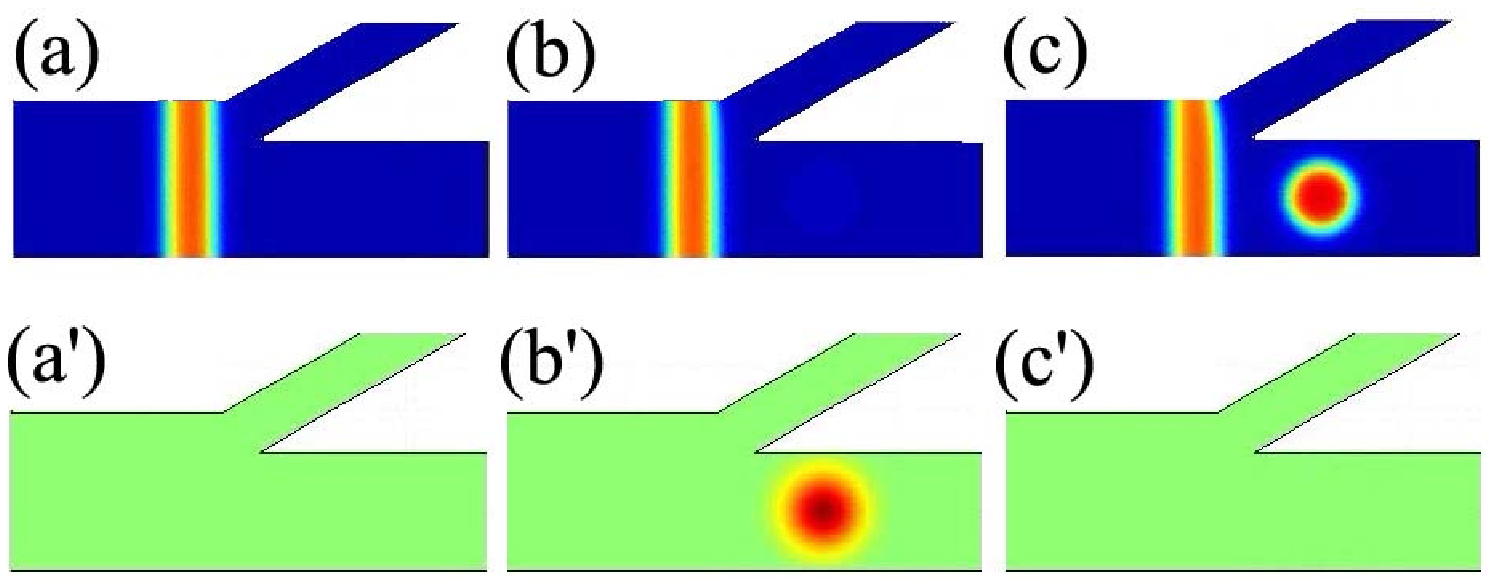}
\includegraphics[width=0.9\columnwidth]{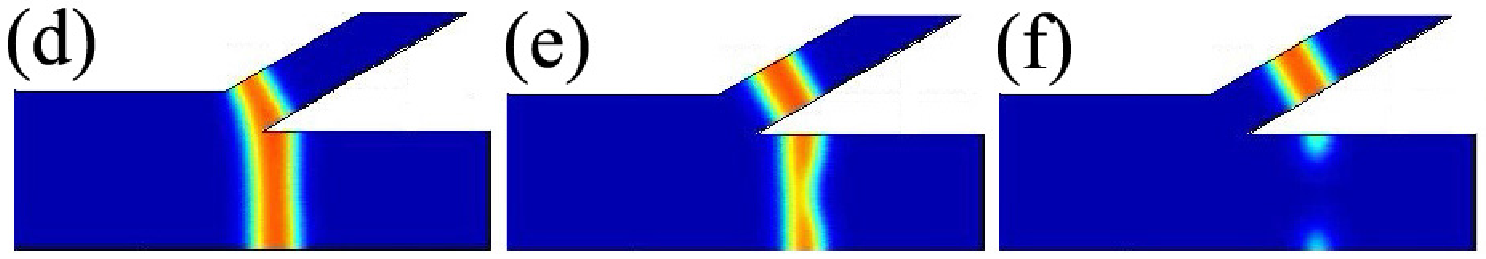}
\caption{(Color online) Using local laser heating to prevent pulse
propagation.
Similar to Fig.~\ref{laser_help}, instantaneous temperature fields
((a')-(c')) are plotted below the corresponding oxygen coverage
((a)-(c)).
The local laser heating is turned on for a total of 0.4s (from
t=1.5s to 1.9s) and centered in the middle of the lower channel
entrance with a maximum temperature increase of $4.5K$.
(a) t=1.0s, (b) t=1.5s, (c) t=2.0s, (d) t=7.0s, (e) t=10.0s, (f)
t=12.0s.
$T=535.5K$, $P_{CO}=4.95\times10^{-5}mbar$,
$P_{O_2}=2.0\times10^{-4}mbar$.} \label{laser_prevent}
\end{figure}

Laser heating can also be used to eliminate a pulse, as shown in
Fig.~\ref{laser_prevent}.
First, we apply local laser heating for a short time to locally
ignite the CO-poisoned catalytic surface far ahead from the pulse
without creating a new pulse (Figs.~\ref{laser_prevent}(b) and
\ref{laser_prevent}(c)).

The local surface slowly reverts to the quenched steady state
after the laser heating is turned off (as in
Fig.~\ref{laser_prevent}(d)).
There exists a refractory period before this area of the surface
can be reignited, because the dynamics of the Pt
phase-reconstruction are slow.
If a propagating pulse passes through this area before the local
surface recovers, the pulse may not propagate through this area
and ``evaporates" (see Figs.~\ref{laser_prevent}(e) and
\ref{laser_prevent}(f)).

\section{Experiments}

\subsection{Experimental setup}
In earlier experiments, propagating chemical waves had to be
visualized in the presence of inert Ti boundaries
\cite{exp1,exp2}.
These measurements were all performed using a
photoemission electron microscope (PEEM), which has a sufficient
lateral resolution and images the changes of the work function due
to the adsorbed species.
To achieve its high resolution, the sample
has to be at a distance of 4 mm to the electron cathode lens,
restricting severely any access to the surface and,
in particular, preventing laser beam addressing.

At first look, ellipso-microscopy for surface imaging (EMSI)
seems to be a more appropriate choice, since it leaves the whole surface
totally open to additional experiments \cite{RAM}.
But EMSI
relies on differences in adsorbed layer thickness.
A microstructured surface features boundaries, laterally limiting
the chemical waves, that are made out of Ti or similar metals and
are several hundred ${\AA}$ tall.
The signals from those boundaries would easily saturate the image,
leaving no contrast to observe the reactive pattern formation.
A different imaging method was therefore necessary, one that
provided good contrast for the observation of the reaction
dynamics, but which was preferably insensitive to the Ti
microstructures.
Additionally a sufficiently
large working distance between the imaging instrument and the
sample was necessary to allow for local addressing of surface
activity by means of focused laser light.

\begin{figure}
\centering
\includegraphics[width=0.9\columnwidth]{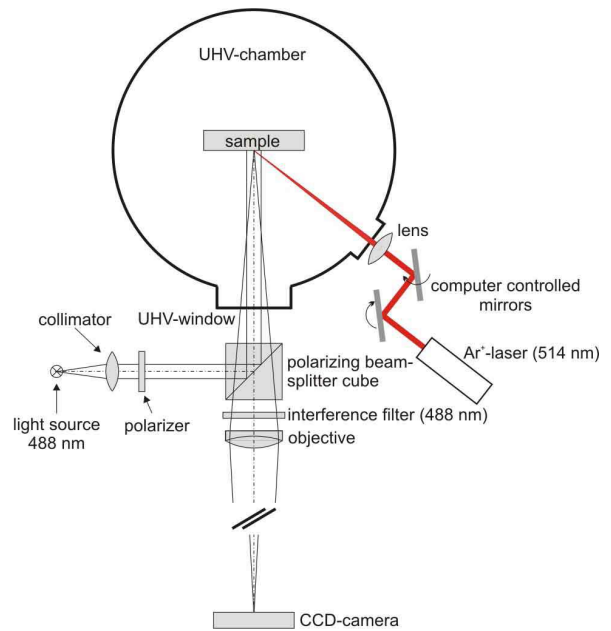}
\caption{Schematic image of the experimental setup (not to scale). For details see
text.} \label{exp_setup}
\end{figure}

Reflection Anisotropy Microscopy (RAM) conveniently combines the
required properties.
RAM has been extensively used to study CO oxidation on platinum.
The first setup basically consisted of a classical
ellipsometric configuration under almost normal incidence
\cite{RAM}.
This setup was later improved by utilizing the
intrinsic properties of a Foster prism and working under exactly
normal incidence \cite{exp3}.
However, the spatial resolution of RAM ($30{\mu}m$) was
insufficient to image the relatively weak and thin reaction pulses
presented in this paper.
In order to improve the spatial resolution significantly, a new
instrument was designed and built. Figure~\ref{exp_setup} shows a
sketch of the experimental setup.

The Pt single crystal is located in a UHV chamber.
As a light source for RAM the $488nm$ line of an $Ar^+$ laser is
used.
The laser
light is collimated and afterwards polarized by means of a
Glan-Thompson prism.
The dielectric film within the polarizing
beam splitter cube reflects about 99.9\% of the incoming
vertically pre-polarized light onto the Pt sample.
During
reflection at the Pt(110) surface the polarization may change
depending on the coverage dependent surface reconstruction
\cite{exp3}.
The light propagates back to the beam splitter.
The
component with retained vertical polarization is reflected back to
the light source.
The other component which is polarized
parallelly, due to interactions with the anisotropic Pt(110)
surface, is transmitted and used for imaging of the crystal.
Twice the light has to pass a UHV window.
This window is specially
designed to minimize stress-induced birefringence.
The spatial resolution of this new setup is $8{\mu}m$.
Thus, also thin pulses
close to the limit of stability can now be observed with RAM.
Through a second UHV window, the light of an additional $Ar^+$
laser ($514nm$) can be focused onto the Pt surface.
The position of the laser spot can be controlled via two
computer-controlled galvanic mirrors.
In order to remove unwanted $514nm$ stray light from the RAM
imaging path, a dielectric filter is used.

\subsection{Experimental results}
\begin{figure}
\centering
\includegraphics[width=0.9\columnwidth]{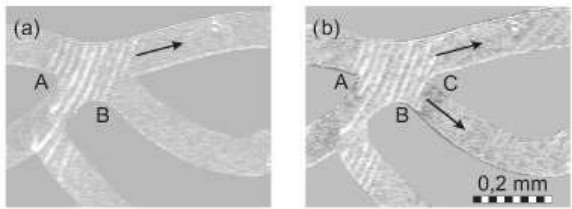} \caption{Time-lapse images
(an interval of 3s between each two snapshots) showing a dying
oxygen-rich pulse ``rescued" by a laser shot centered at $B$.
The arrows indicate the propagation direction of the pulses.
$T= 455K, P_{o_2}= 2\times10^{-4} mbar$, $P_{co}=6.42\times10^{-5}
mbar$.
(a) No laser shot is applied.
The pulse decollates at $B$ and enters the upper right channel.
(b) A laser shot is applied for $50ms$ with a power of $500mW$
when the lower end of the pulse reaches point $B$.
The pulse does not decollate at $B$, and splits into two at $C$
entering both channels.} \label{river_delta}
\end{figure}

\begin{figure}
\centering
\includegraphics[width=0.9\columnwidth]{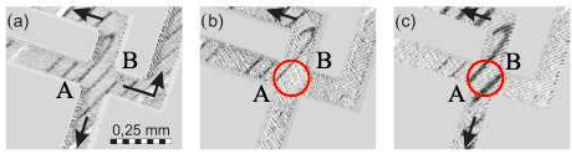}
\caption{Guidance of CO-rich pulses in a labyrinthine structure.
The circles indicate the location where the laser shot is applied.
The laser shot duration is 1s with a power of $230mW$, which
destroys any part of the pulse within the yellow circle.
$T= 451K, P_{o_2}= 2\times10^{-4} mbar$, $P_{co}=5.50\times10^{-5}
mbar$.
(a) No laser shot is applied and the pulse enters all the three
channels.
(b) A laser shot is applied when the lower end of the pulse is at
corner $A$ of the crossing.
The pulse only enters the upper channel.
(c) A laser shot is applied when the upper end of the pulse passes
corner $B$ of the crossing.
The pulse is thus prevented from entering the channel on the
right.} \label{crossroads}
\end{figure}

In the experiments, we were able to guide pulse propagation
through different local laser heating protocols.
We can assist the propagation of  a pulse in the direction
originally prohibited by the boundary curvature-induced
instability through a short duration laser shot
(Fig.~\ref{river_delta}); alternatively, we can apply excessive
heating that actually destroys the pulses (Fig.~\ref{crossroads}).

In Fig.~\ref{river_delta}, an oxygen-rich pulse is first ignited
at the upper left channel and propagates toward the right.
It is able to turn around corner $A$, but it decollates at corner
$B$ unless it is assisted by local laser shot.
After applying a laser shot at the lower end of the pulse at $B$,
the increase of local temperature reduces the CO coverage of the
local CO-poisoned surface, which makes the adsorption of oxygen
easier, so that the pulse is able to pass around the $B$ corner
(Fig.~\ref{river_delta}(b)).

If excessive heating is applied to a propagating CO-rich pulse,
removing enough CO from the local surface, the pulse can be
locally destroyed.
In Fig.~\ref{crossroads}(b), we apply the laser heating for 1s
when the lower end of the pulse reaches corner $A$ of the
crossing.
This erases the lower part of the pulse within the yellow circle
and prevents the pulse from entering the right and lower channels.
In Fig.~\ref{crossroads}(c), we wait until the {\it upper end} of
the pulse passes the $B$ corner  of the crossing before applying
the laser heating.
Only the pulse propagation in the right channel is thus obstructed.

\section{Summary and Conclusions}

Spontaneous pattern formation, and the dynamics of the resulting
patterns in reacting systems, can be controlled at several levels.
Designing the geometry of the catalytic domain, and using a
galvanometer mirror-manipulated laser beam, we have demonstrated
here through experiment and computation, that the propagation of
pulses in complex geometries can be guided, facilitated or
forbidden in real time.
Sudden and intense boundary curvature changes can lead to a
fundamental ``decollation" instability; on the other hand the
laser-induced heating can either assist or prohibit pulse
propagation depending on its intensity and location.

The combination of spatially fine-grained sensing with
spatially fine-grained actuation opens up a wide array
of possibilities in manipulating physical processes in
complex microgeometries.
Here we studied heterogeneous reacting systems; similar tools can
be used to implement spatiotemporal networks in {\it homogeneous}
reacting excitable media (e.g. Ref.~\cite{tinsley}) or
electrochemically reacting systems (e.g. Ref.~\cite{Kiss}).
Beyond chemically reacting systems, such tools are becoming
increasingly useful -and used- in applications as diverse as
droplet formation and mixing in microfluidics~\cite{Ismagilov},
directing cell migration~\cite{Jiang-Whitesides} or manipulating
coherent matter-waves~\cite{Oberthaler}.

{\bf Acknowledgements} This work was partially supported by an
NSF/ITR grant and by AFOSR (IGK, LQ); LQ gratefully acknowledges
the support of a PPPL Fellowship. The authors also wish to
acknowledge the experimental observations of rhomb constrictions
in Michael Pollmann's Thesis at the FHI.


\end{document}